# Relativistic *ab initio* study on the spectroscopic and radiative properties of the lowest states and modeling of the optical cycles for the LiFr molecule

Maksim Shundalau * and Patrizia Lamberti

*Department of Information and Electrical Engineering and Applied Mathematics,*
*Università di Salerno, via Giovanni Paolo II, 84084 Fisciano (SA), Italy*
*\* Corresponding author, e-mail: mshundalau@unisa.it*

**Abstract.** The LiFr diatomic represents a promising candidate for indirect laser cooling that has not yet been investigated not theoretically or experimentally. The potential energy curves of the ground and low-lying excited states of the LiFr heteronuclear alkali metal dimer are calculated using the Fock-space relativistic coupled cluster theory for the first time. A number of properties such as the electronic term energies, equilibrium internuclear distances, transition and permanent dipole moments, sequences of vibrational energies, harmonic vibrational frequencies, Franck–Condon factors, and radiative lifetimes (including bound and free transitions) are predicted. The probabilities of the two-step schemes (optical cycles) for the transfer process of the LiFr molecules from high excited vibrational states to the ground vibronic state are also predicted. The data obtained would be useful for laser cooling and spectral experiments with LiFr molecules.

## 1. Introduction

During the last decades, alkali metal diatomics including LiNa [1], KRb [2], NaK [3], RbCs [4], and others became objects for the effective production of ultracold matter having manifold potential applications such as the creation of a Bose–Einstein condensate, quantum information processing, controlled chemical reactions [5], and research for parity-violation effects [6].

Unlike other alkali metal diatomics, which were well-studied by both theoretical and experimental methods, the francium diatomic compounds are the focus of only a few theoretical works mainly because francium has no stable long-lived isotopes. Derevianko et al [7–9] and Marinescu et al [10] calculated the Lennard–Jones and van der Waals coefficients for Fr dimers [7, 8, 10] and heteronuclear LiFr dimers [9]. Some parameters of the $Fr_2$ ground state were found by Roos et al [11], Lim et al [12], Noro et al [13], Pershina et al [14], and Hill and Peterson [15]. Aymar et al [16] calculated potential energy curves (PECs) and other characteristics of the ground singlet state and first excited triplet state of the francium $Fr_2$ dimer and heteronuclear

RbFr and CsFr dimers. Recently, Jellali and Habli [17] obtained characteristics of the numerous low-lying terms of the $Fr_2$ dimer and $FrCa^+$, $FrSr^+$, and $FrBa^+$ ions. The ground [15, 18] and excited [18] states of the FrF [15] and FrH [15, 18] diatomics were investigated by Hill and Peterson [15] and Souissi et al [18]. The molecular spectroscopic parameters of the lowest terms of the FrCu, FrAg, and FrAu molecules were calculated by Pershina et al [14], Śmiałkowski and Tomza [19], and Kłos et al [20]. The mentioned molecules have an electronic structure similar to heteronuclear alkali metal diatomics. Finally, heteronuclear NaFr and CsFr dimers were recently investigated by Jellali et al [21] and Alyousef et al [22]. Note that nowadays, the francium atom as the heaviest of the alkali metal atoms is considered one of the prospective candidates for searching for the electron electric dipole moment [23–26].

So, previously most efforts have been focused rather on the francium dimer [11–17] and to the best of our knowledge, neither theoretical nor experimental data are available for the lowest electronic states of the LiFr molecule. Lately, we performed the state-of-the-art MRPT (multi-reference perturbation theory) and FS-RCCSD (Fock-space relativistic coupled cluster singles and doubles) calculations for alkali metal and alkaline earth metal-containing diatomics [27–34]. Here we use the latter level of theory expecting to achieve the best results for the LiFr molecule and fill in one of the blanks.

The main goals of the work are: (i) to calculate the potential energy curves of the ground and low-lying excited states of the LiFr diatomic at the high level of theory, (ii) to determine other spectroscopic characteristics of the electronic states (transition dipoles, vibrational energies, Franck–Condon factors, lifetimes, etc.) and, finally, (iii) to find out the possibilities of the effective optical loop realization for the laser cooling of the LiFr molecules.

## 2. Computational details

*Ab initio* relativistic calculations were performed using Kramers unrestricted IH-FS-RCCSD (intermediate Hamiltonian Fock-space relativistic coupled cluster singles and doubles) [35, 36] method. Nowadays, the FS-RCC approach is one of the most successful tools for predicting the electronic structure and properties of molecular compounds containing heavy atoms. It provides the most accurate data on PECs and other characteristics of excited states of small molecular systems. Its theoretical accuracy for the predicted electronic transition energies is evaluated to be less than 100 $cm^{-1}$ for alkaline earth monohalides [33, 37, 38] and even better for some alkali metal diatomics [39, 40].

In terms of the FS-RCC approach, an effective Hamiltonian is defined in a model space, which is constructed from Slater determinants. A reference zero-order wave function (or vacuum state) is a closed-shell determinant, and the operator of the excitation is defined relative to the vacuum state and divided into parts according to the number of valence holes and valence electrons. In the case of an alkali metal diatomic molecule, the vacuum state is the closed-shell doubly positively charged ion (or 0 holes and 0 electrons over vacuum). Thus, the first step of calculations for an alkali metal diatomic molecule in terms of the FS-RCC approach is a solution of the coupled cluster equations for a closed-shell reference dication (or (0,0) Fock sector). The second step is adding an electron and solving the coupled cluster equations for an open-shell positively charged molecular ion (or (0,1) Fock sector). The final step is adding one more electron and solving equations for a closed-shell neutral molecule (or (0,2) Fock sector). To avoid intruder states and convergence difficulties, the intermediate Hamiltonian (IH) formalism [36] was used.

Calculations were started with generating pseudospinors at the HF-SCF level of theory for the closed-shell ground state of LiFr$^{2+}$ ion. The IH-FS-RCCSD calculations in the (0,2) Fock sector (0 holes, 2 electrons, or 2 particles over vacuum) were performed for the LiFr$^{2+}$ ion for 19 low-lying states. The Stuttgart ECPDS78MDFSO fully relativistic large effective core potential (ECP) [41] was used for the francium atom. The ECP [41] replaces 78 chemically inactive core electrons for the francium atom by empirical pseudo-potentials. It also includes the spin-orbit (SO) parameters and takes into account the Breit interactions in the computational scheme (see also [42]). The Gaussian cc-pCVnZ-PP (n = 3, 4, 5), cc-pwCVnZ-PP (n = 3, 4, 5), and aug-cc-pCVnZ-PP (n = 3, 4) [43] basis sets were used for the remaining francium electrons. We also used cc-pCVnZ (n = 3, 4, 5) and cc-pwCVnZ (n = 3, 4, 5) [44] all-electron basis sets for the lithium atom. Since no experimental data are evaluable for a system of electronic states of the LiFr molecule, we compared the calculated energies of molecular states at the dissociation limits (namely, at the internuclear distance of 20.0 Å) with the sum of the NIST experimental energies [45] of separated atoms. Among all trial calculations, the best agreement with the NIST [45] data for dissociation limits was achieved by using the cc-pwCVTZ-PP (Fr) + cc-pwCVTZ (Li) basis sets combination. These basis sets were further used for the calculations.

All nine remaining electrons (namely, eight subvalence, or outer core 6s$^2$6p$^6$ electrons and one valence 7s electron) of the francium atom and all three electrons (1s$^2$2s$^1$) of the lithium atom were included in correlation calculations. The calculations were performed pointwise for the 2.00–20.00 Å internuclear distances by steps of 0.10, 0.05, 0.10, 0.25, 0.50, and 1.00 Å in (2.00–3.00), (3.00–5.00), (5.00–6.00), (6.00–9.00), (9.00–15.00), and (15.00–20.00) Å regions, respectively.

The direct evaluation of electric permanent and transition dipole moments (PDMs and TDMs) does not implement in the FS-RCC method. Therefore, the calculations of the TDMs and PDMs were performed at the MRCI/cc-pwCVTZ level of theory.

All PECs, PDMs, and TDMs calculations were performed using the DIRAC19 quantum chemical package [42]. The electronic energy terms $T_e$ and the equilibrium internuclear distances $R_e$ were obtained using the second-degree polynomial approximation of the *ab initio* PECs near their minima. The calculations of the vibrational state energies and the Franck–Condon factors were performed using the LEVEL program package [46].

## 3. Results and discussion

*3.1. Electronic and vibrational states*

The PECs calculations have been performed for the states belonging to the three lowest dissociation limits: Li(2s) + Fr(7s), Li(2s) + Fr(7p), and Li(2p) + Fr(7s). Within the Hund's case (a), the ground $X^1\Sigma^+$ state and the first triplet $1^3\Sigma^+$ ($a^3\Sigma^+$) state correspond to the first asymptotic limit, the singlet $2^1\Sigma^+$ and $1^1\Pi$ states and the triplet $1^3\Pi$ and $2^3\Sigma^+$ states correspond to the second one, and the singlet $3^1\Sigma^+$ and $2^1\Pi$ states and the triplet $2^3\Pi$ and $3^3\Sigma^+$ states correspond to the third one. Since the spin-orbit coupling for the francium atom is quite large [45], it should be taken into account. Within the Hund's case (c), the PECs are labeled by $(n)\Omega^\sigma$ notation, where $\Omega$ is the absolute value of the projection of the total electronic angular momentum on the LiFr axis and $\sigma = \pm$ shows the even (+) or odd (−) reflection symmetry of the electron eigenfunction concerning a plane, which contains the internuclear axis if $\Omega = 0$. Numbers $n = 1, 2, \ldots$ label states of the same $\Omega^\sigma$ value ordered by increasing energy. In this case, the $X^1\Sigma^+$, $2^1\Sigma^+$ and $3^1\Sigma^+$ terms correspond to the $(X)0^+$, $(2)0^+$ and $(4)0^+$ terms, respectively; the $1^1\Pi$ and $2^1\Pi$ terms correspond to the $(4)1$ and $(7)1$ terms, respectively; the $1^3\Sigma^+$, $2^3\Sigma^+$ and $3^3\Sigma^+$ terms split into the $(1)0^-$ and $(1)1$, $(2)0^-$ and $(2)1$ and $(4)0^-$ and $(5)1$ terms, respectively; the $1^3\Pi$ and $2^3\Pi$ terms split into the $(3)0^+$, $(3)0^-$, $(3)1$ and $(1)2$ and $(5)0^+$, $(5)0^-$, $(6)1$ and $(2)2$ terms, respectively.

All calculated PECs versus internuclear distances are shown in Fig. 1 and are also presented in the supplementary material. For clarity, the PECs for $\Omega = 0^\pm$, 1, and 2 are also shown in Figs. S1–S4 (see supplementary material) separately.

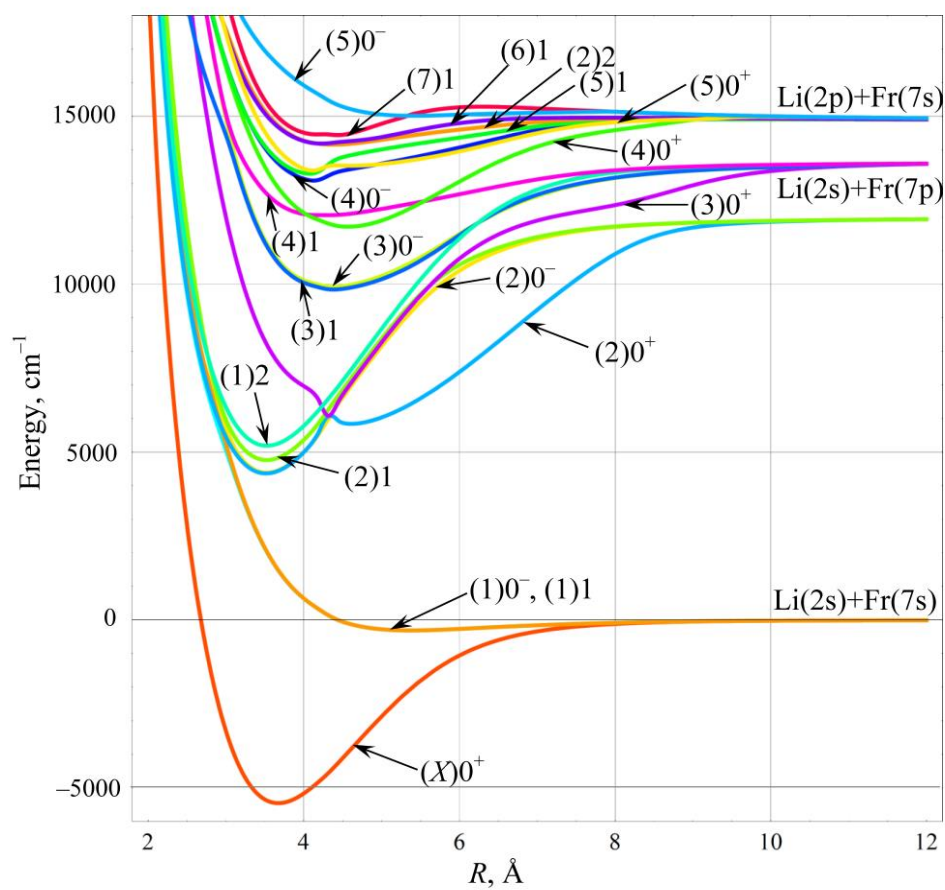

Fig. 1. The calculated PECs for the low-lying states of the LiFr molecule.

Lithium has two stable isotopes ($^6$Li and $^7$Li) with an abundance of the latter between 92.2% and 98.1%. Francium has no stable isotopes; the longest-lived isotope of francium is $^{223}$Fr with a half-life of about 22 minutes. We calculated the vibrational energies for the $^7$Li$^{223}$Fr molecule for nineteen low-lying PECs and then obtained the harmonic vibrational frequencies $\omega_e$ and other molecular spectroscopic constants for these electronic states (see Table 1).

Table 1. Molecular spectroscopic constants of the $^7$Li$^{223}$Fr molecule.

| Limit | Limit energy, cm$^{-1}$ | | Term notation | $T_e$, cm$^{-1}$ | $R_e$, Å | $D_e$, cm$^{-1}$ | $\omega_e$, cm$^{-1}$ | $\omega_e\chi_e$, cm$^{-1}$ | $B_e$, cm$^{-1}$ |
|---|---|---|---|---|---|---|---|---|---|
| | Exp. [a] | Theor. | | | | | | | |
| Li(2s)+Fr(7s) | 0.0 | 0.0 | (X)0$^+$ | 0.0 | 3.673 | 5457.6 | 180.9 | 1.355 | 0.1831 |
| | | | (1)0$^-$ | 5143.5 | 5.378 | 314.1 | 42.8 | 1.608 | 0.0843 |
| | | | (1)1 | 5143.8 | 5.380 | 313.8 | 42.7 | 1.609 | 0.0842 |
| Li(2s)+Fr(7p) | 12237.4 | 11946.9 | (2)0$^+$ | 9821.4 [b] / 11298.9 [c] | 3.513 [b] / 4.604 [c] | 7583.2 | 180.2 | 0.480 | 0.2004 |
| | | | (2)0$^-$ | 9845.0 | 3.521 | 7559.5 | 183.4 | 0.306 | 0.1993 |
| | | | (2)1 | 10219.3 | 3.522 | 7185.2 | 181.3 | 0.574 | 0.1992 |
| | 13924.0 | 13606.2 | (3)0$^+$ | 11539.3 | 4.317 | 7524.5 | — | — | 0.1329 |
| | | | (3)0$^-$ | 15376.1 | 4.380 | 3687.7 | — | — | 0.1284 |
| | | | (3)1 | 15298.8 | 4.381 | 3765.0 | — | — | 0.1281 |
| | | | (1)2 | 10647.2 | 3.524 | 8416.7 | 180.2 | 0.718 | 0.1991 |
| | | | (4)1 | 17513.0 | 4.321 | 1550.9 | — | — | 0.1350 |
| Li(2p)+Fr(7s) | 14903.7 | 14898.1 | (4)0$^+$ | 17168.0 | 4.564 | 3187.7 | 112.1 | 1.869 | 0.1182 |
| | | | (4)0$^-$ | 18543.6 | 4.126 | 1812.1 | — | — | 0.1467 |
| | | | (5)1 | 18749.0 | 4.073 | 1606.8 | — | — | 0.1516 |
| | | | (2)2 | 19613.7 | 4.382 | 742.3 | 79.6 | 2.430 | 0.1275 |
| | | | (6)1 | 19646.9 | 4.222 | 709.1 | 88.1 | 4.112 | 0.1355 |
| | 14904.0 | 14903.3 | (5)0$^+$ | 18985.9 | 4.602 | 1375.0 | — | — | 0.1448 |
| | | | (7)1 | 19897.6 | 4.436 | 463.3 | — | — | 0.1323 |
| | | | (5)0$^-$ | 20470.5 | 5.400 | — | — | — | 0.0827 |

Notes:
[a] Ref. [45];
[b] First minimum;
[c] Second minimum.

Most of the lowest terms have rather regular character, excluding the (2)0$^+$ and (3)0$^+$ ones, which repulse each other near 4.35 Å to avoid crossing. As a result, the (2)0$^+$ term has two minima at 3.51 and 4.60 Å, and in its turn the (3)0$^+$ term has an extremely specific shape with a narrow ledge in the minimum region. As a consequence, the sequence of its vibrational levels cannot be described in the terms of a two-parameter ($\omega_e$ and $\omega_e\chi_e$) anharmonic oscillator (see below). A few other higher excited terms (for example, (3)0$^-$, (3)1, and others) also have peculiarities due to the same reasons. Therefore, for them, it is also impossible to determine harmonic frequency and anharmonicity constant (see Table 1). Fig. 2 shows calculated vibrational energies $E_v$ versus vibrational quantum number $v$. The insert demonstrates the mentioned dependence for the lowest vibrational levels of the (3)0$^+$ term.

In general, the system of the lowest PECs for the LiFr molecule is typical for alkali metal dimers. Some experimental (if they are available) and theoretical molecular spectroscopic parameters for the ground $X^1\Sigma_g^+$ ($X^1\Sigma^+$) and first triplet $a^3\Sigma_u^+$ ($a^3\Sigma^+$) states for the alkali metal dimers containing lithium or/and francium atoms are listed in Table 2. There are trends to monotone decreasing the dissociation energy and increasing equilibrium internuclear distance for the ground state in the series LiM and MFr (where M = Li, Na, K, Rb, Cs, and Fr) diatomics. Regarding the first triplet state, it is worth mentioning that its dissociation energy is rather independent of the second atom in the LiM and MFr series. At the same time, the equilibrium internuclear distance for the first triplet state monotonically increases. Note also that the spin-orbit splitting of the $a^3\Sigma^+$ state into (1)0$^-$ and (1)1 PECs for the LiFr molecule is less than 1 cm$^{-1}$ in the region of bound states.

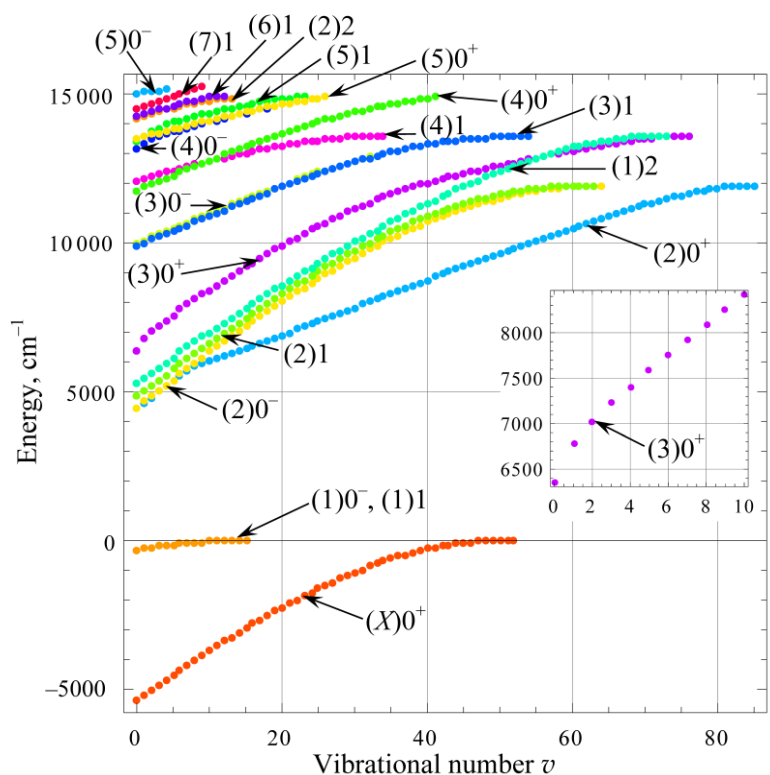

Fig. 2. The vibrational energies $E_v$ for the low-lying states of the $^7$Li$^{223}$Fr molecule.

Table 2. Selected molecular spectroscopic constants of lithium and francium diatomics.

| Molecule | Ground state $X^1\Sigma_g^+/X^1\Sigma^+$ | | First triplet state $a^3\Sigma_u^+/a^3\Sigma^+$ | | Experimental/ theoretical | References, approximation |
|---|---|---|---|---|---|---|
| | $D_e$, cm$^{-1}$ | $R_e$, Å | $D_e$, cm$^{-1}$ | $R_e$, Å | | |
| Li$_2$ | 8517 | 2.673 | 333 | 4.171 | experimental | [47, 48] |
|        | 8466 | 2.677 | 334 | 4.169 | theoretical  | [39], FS-CCSD [a]        |
| LiNa   | 7105 | 2.885 | 230 | 4.701 | experimental | [49, 50]                 |
|        | 7089 | 2.896 | 244 | 4.677 | theoretical  | [40], FS-CCSD [a]        |
| LiK    | 6216 | 3.318 | 287 | 4.99  | experimental | [51, 52]                 |
|        | 6208 | 3.296 | 376 | 4.966 | theoretical  | [53], FS-CCSD [b]        |
| LiRb   | 5927 | 3.466 | 277 | 5.140 | experimental | [54]                     |
|        | 5924 | 3.45  | 291 | 5.1   | theoretical  | [55], FS-CCSD [c]        |
| LiCs   | 5875 | 3.668 | 309 | 5.247 | experimental | [56]                     |
|        | 5872 | 3.65  | 295 | 5.2   | theoretical  | [55], FS-CCSD [c]        |
| LiFr   | 5458 | 3.673 | 314 | 5.380 | theoretical  | This work, FS-CCSD [c]   |
| NaFr   | 4506 | 3.77  | 164 | 6.27  | theoretical  | [21], Full CI [d]        |
| KFr    | —    | —     | —   | —     | —            | —                        |
| RbFr   | 3654/3690 | 4.3 | 207/210 | 6.3 | theoretical | [16], Full CI [d]    |
| CsFr   | 3553/3576 | 4.52 | 210/218 | 6.53 | theoretical | [16], Full CI [d]  |
|        | 3518 | 4.53  | 312 | 6.58  | theoretical  | [22], Full CI [d]        |
|        | 3283 | 4.665 | —   | —     | theoretical  | [11], CASPT2 [a]         |
|        | 3476 | 4.593 | —   | —     | theoretical  | [12], CCSD(T) [c]        |
|        | 3498/3576 | 4.47 | 188/201 | 6.61 | theoretical | [16], Full CI [d]  |
| $Fr_2$ | 3250 | 4.698 | —   | —     | theoretical  | [13], CCSD(T) [e]        |
|        | 3517 | 4.610 | —   | —     | theoretical  | [14], DFT [f]            |
|        | 3088 | 5.064 | —   | —     | theoretical  | [15], CCSD(T) [h]        |
|        | 3345 | 4.49  | 179 | 6.61  | theoretical  | [17], Full CI [d]        |

Notes:
[a] ANO-RCC all-electron basis set;
[b] Sadlej's polarized TZ all-electron basis set;
[c] 9-valence-electron PP basis set;
[d] two correlated electrons, one-valence-electron large core PP basis set;
[e] all-electron NOSeC-CV-qzp basis set;
[f] 4c relativistic DFT, B88/P86 functionals, all-electron 4c-basis set;
[h] CBS limit for 9-valence-electron cc-pVnZ-PP basis sets.

*3.2. Franck–Condon factors and radiative properties*

It is assumed that the given optical cycle involves some of the lowest $(X)0^+$, $(1)0^-$, $(1)1$, $(2)0^+$, $(3)0^+$, and $(2)1$ terms (see below). For the prediction of the STIRAP parameters we calculated the FCF distributions for the transitions from the upper $(2)0^+$, $(3)0^+$, and $(2)1$ states to the ground $(X)0^+$ state (Fig. 3) and from the upper $(2)0^+$, $(3)0^+$, and $(2)1$ states to the components of the triplet $a^3\Sigma^+$ state ($(1)0^-$ and $(1)1$) (Figs. 4 and 5). The specific shape of the $(2)0^+$ and $(3)0^+$ PECs and the discrepancy between their minima and the minimum of the ground state leads to the presence of several sequences in the FCF distributions (Fig. 3, *a*, *c*), which is likely to be reflected in the complex and irregular nature of the intensity distributions in the emission bands of the $(2)0^+ \rightarrow (X)0^+$ and $(3)0^+ \rightarrow (X)0^+$ transitions.

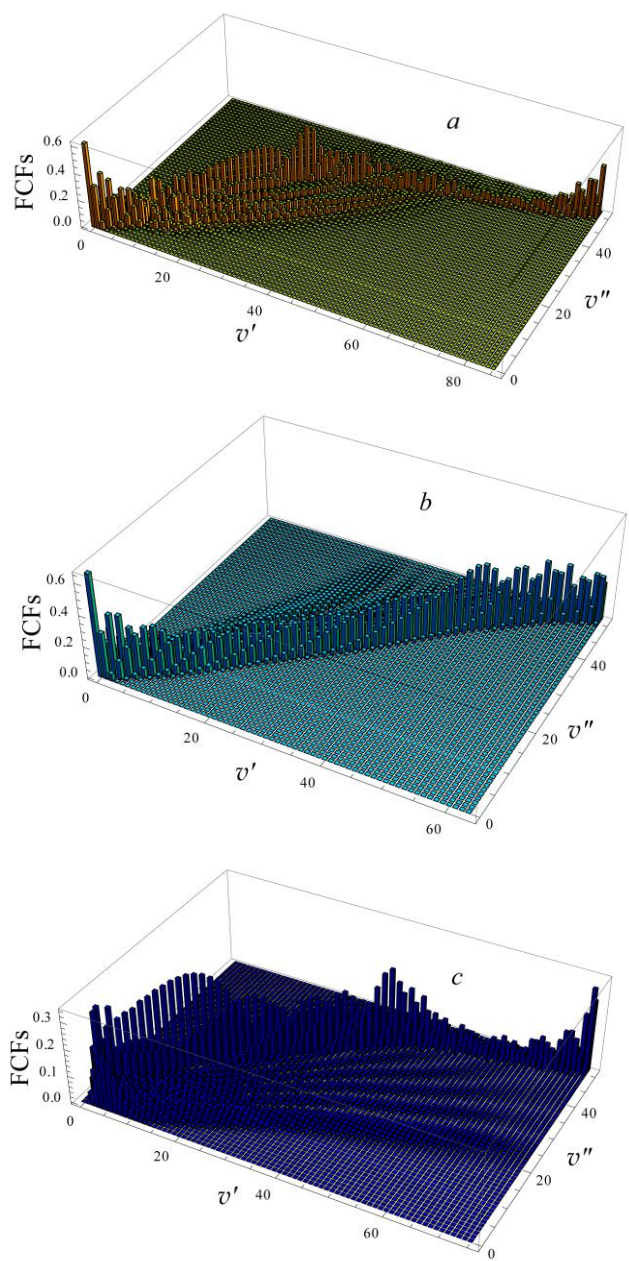

Fig. 3. The calculated FCFs for the $(2)0^+ \to (X)0^+$ (*a*), $(2)1 \to (X)0^+$ (*b*), and $(3)0^+ \to (X)0^+$ (*c*) vibronic transitions of the LiFr molecule.

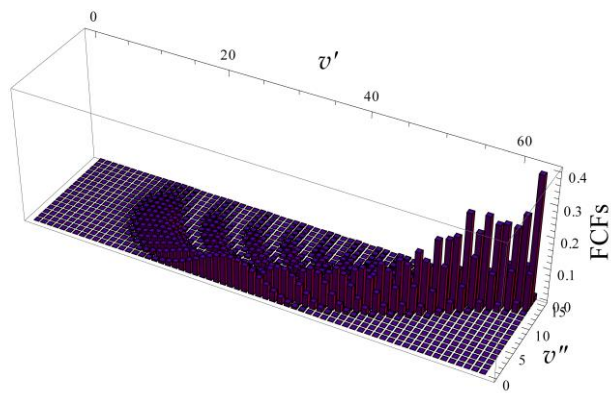

Fig. 4. The calculated FCFs for the $(2)1 \rightarrow (1)0^-$ vibronic transitions of the LiFr molecule.

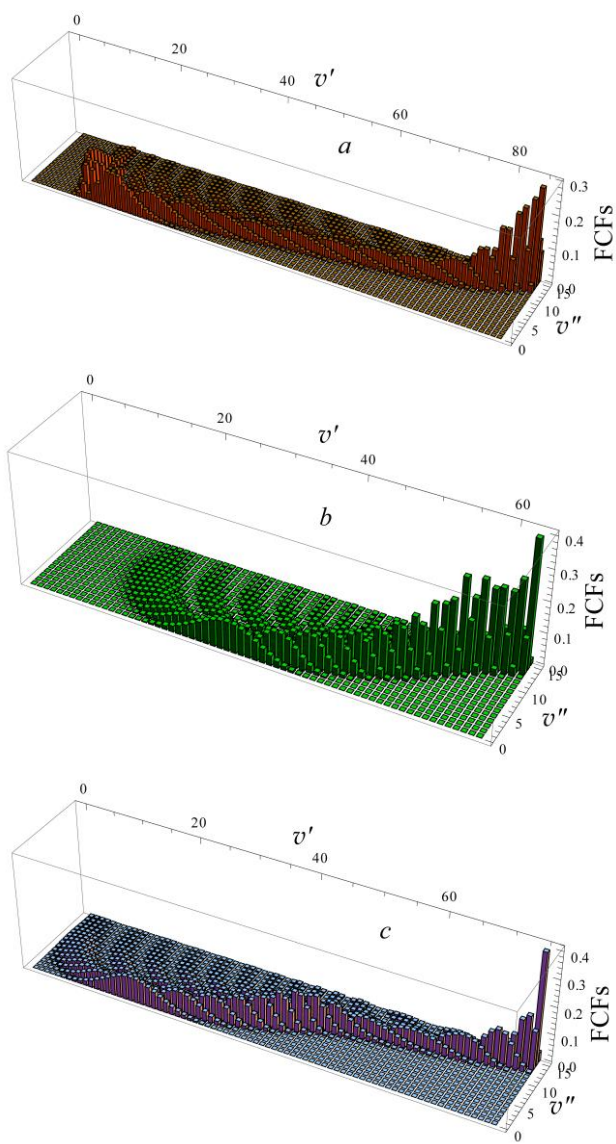

Fig. 5. The calculated FCFs for the $(2)0^+ \rightarrow (1)1$ (*a*), $(2)1 \rightarrow (1)1$ (*b*), and $(3)0^+ \rightarrow (1)1$ (*c*) vibronic transitions of the LiFr molecule.

Calculated permanent dipole moments (PDMs) and transition dipole moments (TDMs) as the functions of internuclear distances for mentioned lowest six terms are shown in Figs. 6 and 7.

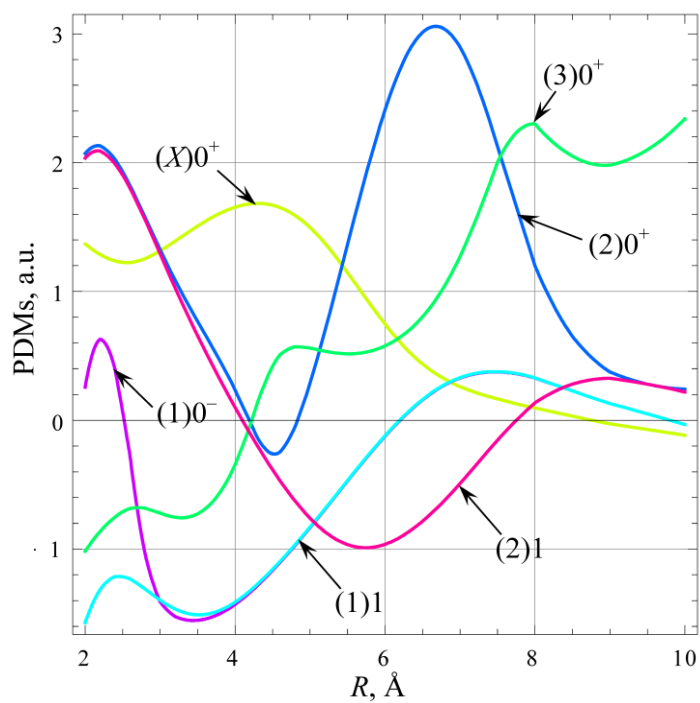

Fig. 6. The calculated permanent dipole moments of the LiFr molecule.

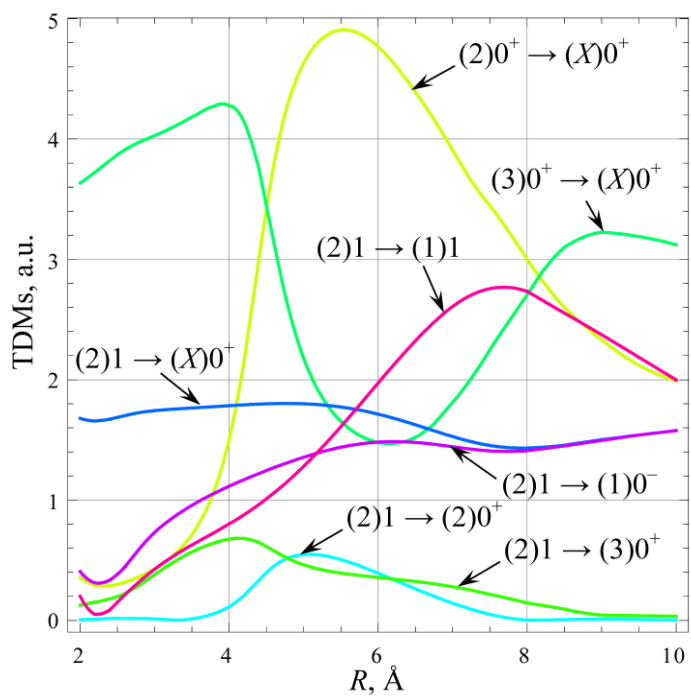

Fig. 7. The calculated transition dipole moments of the LiFr molecule.

The lifetimes of vibrational levels were evaluated based on the calculated TDMs, FCFs, and vibrational levels. The radiative lifetime $\tau_{v'}$ (in seconds) of the vibrational state for bound → bound (b → b) transitions is estimated by the following equation:

$$\tau_{v'}(b \to b) = \frac{4.936 \times 10^5}{|TDM|^2 \sum_{v''} f_{v'v''} (\Delta E_{v'v''})^3},$$

where $|TDM|^2$ is the sum of the squares of absolute values of transition dipole moment between upper and lower electronic states, including longitudinal and two transverse components (in a.u.); $f_{v'v''}$ is the FCF between $v'$ (upper) and $v''$ (lower) vibrational states; $\Delta E_{v'v''}$ is the energy difference between $v'$ and $v''$ (in cm$^{-1}$).

The Einstein coefficient $A_{v'}(b \to f)$ for spontaneous emission arising from bound → free (b → f) transitions into the $(X)0^+$ continuum should be added to obtain the total $A_{v'}$ coefficient and corresponding lifetime [57]:

$$\tau_{v'}(\text{total}) = \frac{1}{A_{v'}(b \to b) + A_{v'}(b \to f)}.$$

The Einstein coefficient $A_{v'}(b \to f)$ is estimated by the equation:

$$A_{v'}(b \to f) = 2.026 \times 10^{-6} |TDM'|^2 \left(1 - \sum_{v''} f_{v'v''}\right) (\Delta E_{v',\text{continuum}})^3,$$

where TDM′ is the TDM at the outer classical turning point of level $v'$; $\Delta E_{v',\text{continuum}}$ is the energy difference between the level $v'$ and the asymptote of the ground term.

The results of calculations of lifetimes of the $(2)0^+$, $(2)1$, and $(3)0^+$ vibrational levels without (sequences 1) and with the b → f term (sequences 2) are presented in Fig. 8. It is worth mentioning that the b → f contribution cannot be neglected, since due to the influence of the $(1 - \sum_{v''} f_{v'v''})$ factor (or the FCF between the level $v'$ and all energies in the continuum) the $A_{v'}(b \to f)$ term increases as the vibrational number increases. For the lowest vibrational levels, the contribution of the $A_{v'}(b \to f)$ term is vanishingly small, but it plays a crucial role for highly excited levels. Taking this into account, the mean lifetimes for vibrational levels of the $(2)0^+$, $(2)1$, and $(3)0^+$ states are 300–130 ns, 150–90 ns, and 20–10 ns, respectively. Including the additional decay channels into the components of the triplet $a^3\Sigma^+$ state ($(2)0^+ \to (1)1$, $(2)1 \to (1)1$ and $(1)0^-$, $(3)0^+ \to (1)1$), the mean lifetimes of the $(2)0^+$ and $(2)1$ states decrease to 245–95 ns and 140–60 ns; the mean lifetimes of the $(3)0^+$ states almost do not change (sequences 3 in Fig. 8).

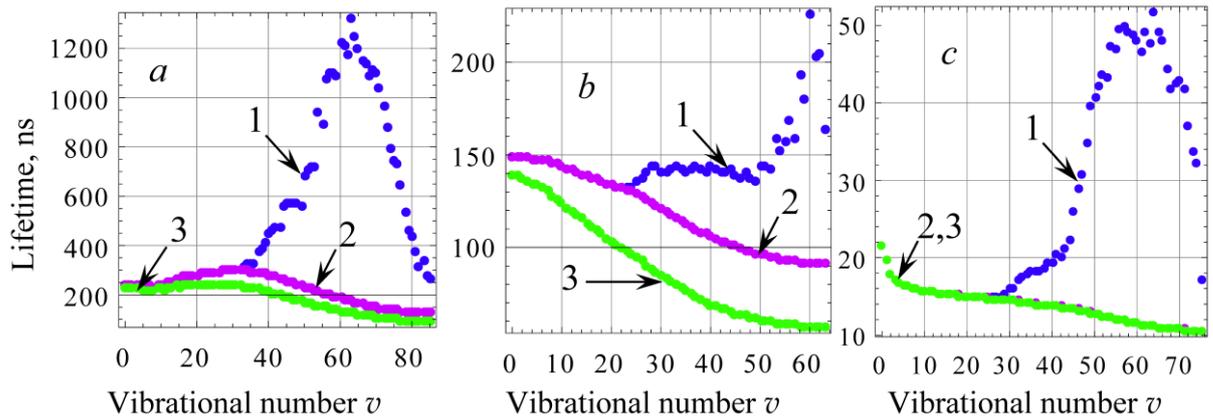

Fig. 8. The radiative lifetimes of the $(2)0^+$ (*a*), $(2)1$ (*b*), and $(3)0^+$ (*c*) vibrational levels

### 3.3. Optical cycles

The production of polar dimers of alkali metals in the ground ("absolute") rovibronic $X^1\Sigma^+(v=0, J=0)$ state can be based on the process of Stimulated Raman Adiabatic Passage (STIRAP, see, for example, [58]). Briefly, the idea of a two-stage optical cycle is as follows. Due to photoassociation, diatomic molecules are formed from pre-cooled atoms with internuclear distances far exceeding the equilibrium value or, in other words, in highly excited (near the dissociation limit) vibrational states of one of the lowest electronic states. Usually, this initial $E_{\text{init}}$ state is the first triplet $a^3\Sigma^+$ state. Due to the low initial translational temperature of atoms, the weakly bound dimers are in rotational states with small values of rotational quantum number *J*. Further, weakly bound dimers with a low rotational temperature, which are in highly excited vibrational states of the initial electronic state $E_{\text{init}}(v'', J'' = 0)$, as a result of laser pumping (channel 1 in Fig. 9) are transferred to the vibrational-rotational states of one of the more highly excited electronic states (intermediate state) $E_{\text{im}}(v', J' = 1)$. The pump laser wavelength can be adjusted to avoid populating rovibronic states $E_{\text{im}}(v', J' \neq 1)$ that are not involved in the scheme. The second step of the optical cycle is realized as a result of stimulated radiative transitions (dumping) of molecules from the excited rovibronic states $E_{\text{im}}(v', J' = 1)$ into the "absolute" rovibronic $X^1\Sigma^+(v'' = 0, J'' = 0)$ state (channel 2 in Fig. 9). A pulse from a second laser phase-synchronized with the pump laser is used to produce coherent pumping and dumping and to avoid undesired spontaneous emission and branching. So, for the high efficiency of the process, it is necessary to choose a suitable cycle scheme $E_{\text{init}}(v'', J'' = 0) \rightarrow E_{\text{im}}(v', J' = 1) \rightarrow X^1\Sigma^+(v'' = 0, J'' = 0)$ so that its probability would be the highest.

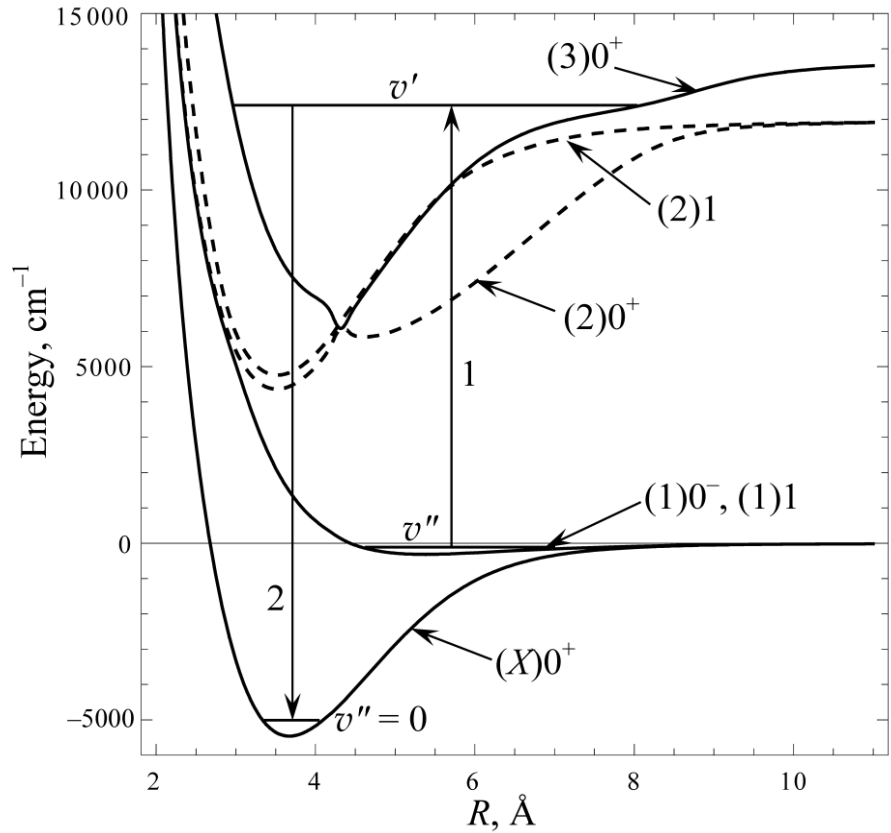

Fig. 9. The optical cycle scheme involving initial $(1)1(v'')$, intermediate $(3)0^+(v')$, and final $(X)0^+(v''=0)$ states.

Thus, to construct an optical cycle, it is first necessary to choose the initial vibrational states whose energies are close to the dissociation limit of the initial $E_{init}$ term that belonged to the first dissociation limit (see Fig. 1). For alkali metal dimers, the initial $E_{init}$ term is usually the first triplet $a^3\Sigma^+$ state or its $(1)0^-$ and $(1)1$ components. In addition, one should also consider the option of excitation from weakly bound vibrational states of the ground $(X)0^+$ term. For example, such a channel is considered for ultracold KCs molecule production [58] and implemented for laser cooling of CsYb dimers [59–61]. Further, among the lower excited terms, it is necessary to choose such intermediate ones that are connected with the ground $(X)0^+$ state by an allowed electric dipole transition. For the LiFr molecule, there are three suitable intermediate terms, namely $(2)0^+$, $(2)1$, and $(3)0^+$. Taking into account the $\Omega$ selection rules ($0^+ \leftrightarrow 0^+$, $0^- \leftrightarrow 0^-$, $1 \leftrightarrow 1$, $0^+ \leftrightarrow 1$, $0^- \leftrightarrow 1$), the following options of optical cycles turn out to be possible:

$(X)0^+ \rightarrow (2)0^+ \rightarrow (X)0^+$,
$(X)0^+ \rightarrow (2)1 \rightarrow (X)0^+$,
$(X)0^+ \rightarrow (3)0^+ \rightarrow (X)0^+$,
$(1)0^- \rightarrow (2)1 \rightarrow (X)0^+$,
$(1)1 \rightarrow (2)0^+ \rightarrow (X)0^+$,
$(1)1 \rightarrow (2)1 \rightarrow (X)0^+$,
$(1)1 \rightarrow (3)0^+ \rightarrow (X)0^+$.

The total probability $P$ of the optical cycle is proportional to the product of the probability $P_{E_{\text{init}} \to E_{\text{im}}}^{v'',J'' \to v',J'}$ of pumping channel $E_{\text{init}}(v'',J'') \to E_{\text{im}}(v',J')$ and the probability $P_{E_{\text{im}} \to X}^{v',J' \to v''=0,J''=0}$ of dumping into the ground state $E_{\text{im}}(v',J') \to X^1\Sigma^+(v''=0, J''=0)$ (here the transitions with selection rules $\Delta J = \pm 1$ are considered) [5, 58, 62, 63]:

$$P \sim P_{E_{\text{init}} \to E_{\text{im}}}^{v'',J'' \to v',J'} P_{E_{\text{im}} \to X}^{v',J' \to v''=0,J''=0},$$

where $P_{E' \leftrightarrow E''}^{v',J' \leftrightarrow v'',J''} \sim \left| E_{E'}^{v',J'} - E_{E''}^{v'',J''} \right| S_{E' \leftrightarrow E''} f_{v'v''} S_{J'J''}$; $E_{E'}^{v',J'}$ and $E_{E''}^{v'',J''}$ are energies of rovibronic states; $S_{E' \leftrightarrow E''}$ is line strength of the electronic transition; $f_{v'v''}$ is FCF; $S_{J'J''}$ is Hönl–London factor [64]. Since the molecules formed as a result of photoassociation have a low rotational temperature, the effect of rotational degrees of freedom on the total probabilities turns out to be insignificant (the Hönl–London factors are equal to unity, the energies of rotational sublevels are small due to the smallness of the rotational constant) and are not considered further.

The calculated normalized probabilities of the optical cycles involving ground $(X)0^+$ term and components of the first triplet $a^3\Sigma^+$ term as initial $E_{\text{init}}$ states are presented in Fig. S5 (see supplementary material) and Figs. 10 and 11, respectively. Each $P$ value with coordinates $(v'', v')$ in the diagrams determines the probability of pumping from the initial vibronic state $E_{\text{init}}(v'')$ into the intermediate state $E_{\text{im}}(v')$ and dumping into the final "absolute" state $(X)0^+(v''=0)$.

These results show that highly excited vibrational states of the ground $(X)0^+$ term do not seem promising for an effective optical cycle (Fig. S5). In this case, the probabilities differ from zero only for transitions from low-lying vibrational states far from the dissociation limit, regardless of the intermediate term used.

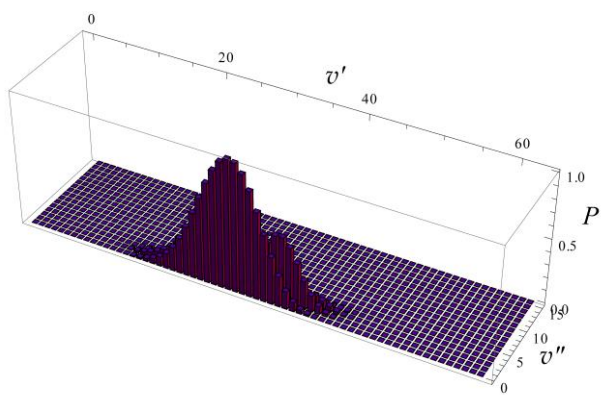

Fig. 10. Probabilities of the $(1)0^-(v'', J''=0) \to (2)1(v', J'=1) \to (X)0^+(v''=0, J''=0)$ optical cycle.

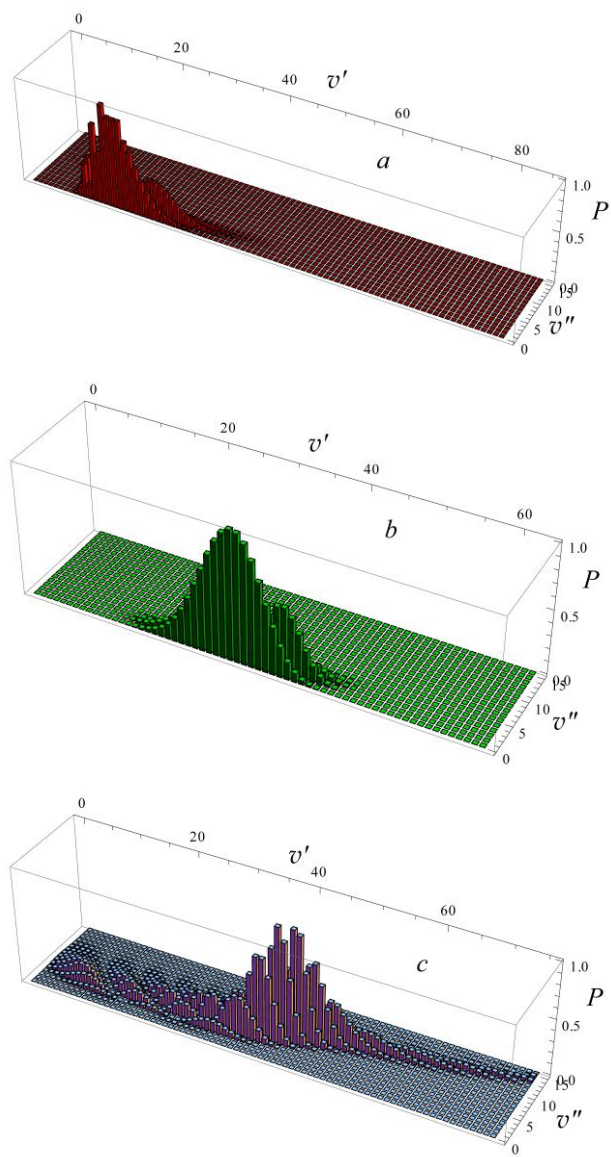

Fig. 11. Probabilities of the (1)1($v''$, $J'' = 0$) → (2)0$^+$($v'$, $J' = 1$) → (X)0$^+$($v'' = 0$, $J'' = 0$) (*a*),
(1)1($v''$, $J'' = 0$) → (2)1($v'$, $J' = 1$) → (X)0$^+$($v'' = 0$, $J'' = 0$) (*b*),
and (1)1($v''$, $J'' = 0$) → (3)0$^+$($v'$, $J' = 1$) → (X)0$^+$($v'' = 0$, $J'' = 0$) (*c*) optical cycles.

The probability distributions for the (1)0$^-$($v''$) → (2)1($v'$) → (X)0$^+$($v'' = 0$) (Fig. 10) and (1)1($v''$) → (2)1($v'$) → (X)0$^+$($v'' = 0$) (Fig. 11, *b*) cycles are similar due to the small difference in (1)0$^-$ and (1)1 PECs. Wherein, the first of them seems to be preferable due to the larger value of the transition moment. Realization of the optical cycle using the initial (1)0$^-$ term needs pumping from the vibronic (1)0$^-$($v'' = 0$, 1) states into the intermediate (2)1($v' = 22…38$) states. However, the vibronic (1)0$^-$($v'' = 0$, 1) states are not close enough to the dissociation limit (290–250 cm$^{-1}$) to provide effective photoassociation, and this imposes restrictions on the implementation of optical cycles. The situation is similar for the (1)1($v''$) → (2)0$^+$($v'$) → (X)0$^+$($v'' = 0$) channel (Fig. 11, *a*).

The most efficient optical cycle for the LiFr molecule can be realized on the (1)1($v''$) → (3)0$^+$($v'$) → (X)0$^+$($v'' = 0$) transitions (Fig. 11, *c*). In this case, the greatest probabilities correspond to pumping from the initial $(1)1(v'' = 4…10)$ vibronic states, which are located 120–30 cm$^{-1}$ below the dissociation limit, into the intermediate $(3)0^+(v' = 30…50)$ states. The $(1)1(v'' = 5) \rightarrow (3)0^+(v' = 39) \rightarrow (X)0^+(v'' = 0)$ cycle has the maximal probability. The normalized probabilities for the $(1)1(v'' = 6) \rightarrow (3)0^+(v' = 41) \rightarrow (X)0^+(v'' = 0)$ and $(1)1(v'' = 6) \rightarrow (3)0^+(v' = 42) \rightarrow (X)0^+(v'' = 0)$ cycles are 0.98 and 0.94, respectively. Probabilities of the optical cycles involving the closest to the dissociation limit initial states ($v'' = -2, -3, -4$) are significantly lower. So, the normalized probabilities for the $(1)1(v'' = -2) \rightarrow (3)0^+(v' = 74) \rightarrow (X)0^+(v'' = 0)$, $(1)1(v'' = -3) \rightarrow (3)0^+(v' = 70) \rightarrow (X)0^+(v'' = 0)$, and $(1)1(v'' = -4) \rightarrow (3)0^+(v' = 66) \rightarrow (X)0^+(v'' = 0)$ are 0.028, 0.040, and 0.053, respectively. These initial states lie 8.5, 13.5, and 20.7 cm$^{-1}$ below the dissociation limit, and it allows the process of photoassociation of atomic pairs.

## 4. Conclusion

For the first time, a theoretical investigation of the lowest terms of the alkali metal LiFr dimer is carried out at the high level of theory (IH-FS-RCCSD). The molecular spectroscopic parameters of the terms belonging to the three lowest dissociation limits, as well as the radiative properties of the vibronic states are predicted. The system of the lowest PECs for the LiFr molecule is typical for alkali metal dimers, and calculated parameters of the LiFr molecule lie in the trends for the series of lithium- and francium-containing diatomics. It was shown that the bound → free transition should be taken into account for the correct evaluation of the radiative lifetimes of the highly excited vibrational levels. Probabilities of optical cycles for the efficient transition of the polar LiFr diatomics into the absolute rovibronic state involving different initial and intermediate excited electronic states are analyzed. It is shown that most efficient two-step laser scheme for the LiFr molecule can be realized on the $(1)1(v'') \rightarrow (3)0^+(v') \rightarrow (X)0^+(v'' = 0)$ transitions.

## Declaration of competing interests

The authors declare that they have no known competing financial interests or personal relationships that could have appeared to influence the work reported in this paper.

## CRediT authorship contribution statement

**Maksim Shundalau:** Conceptualization, Formal analysis, Investigation, Writing – original draft, Visualization. **Patrizia Lamberti:** Writing – original draft, Supervision, Project administration.

## Acknowledgements

This work was partly supported by Horizon 2020 RISE DiSeTCom Project (GA 823728) and University of Salerno funding.

**Supplementary materials**

Supplementary material associated with this manuscript can be found in LiFr-SM.docx file.